# CAPTCHA Based on Human Cognitive Factor

Mohammad Jabed Morshed Chowdhury
Department of Computer Science and Engineering
Daffodil International University
Dhaka, Bangladesh

Narayan Ranjan Chakraborty
Department of Computer Science and Engineering
Daffodil International University
Dhaka, Bangladesh

*Abstract*—A CAPTCHA (Completely Automated Public Turing test to tell Computers and Humans Apart) is an automatic security mechanism used to determine whether the user is a human or a malicious computer program. It is a program that generates and grades tests that are human solvable, but intends to be beyond the capabilities of current computer programs. CAPTCHA should be designed to be very easy for humans but very hard for machines. Unfortunately, the existing CAPTCHA systems while trying to maximize the difficulty for automated programs to pass tests by increasing distortion or noise have consequently, made it also very difficult for potential users. To address the issue, this paper addresses an alternative form of CAPTCHA that provides a variety of questions from mathematical, logical and general problems which only human can understand and answer correctly in a given time. The proposed framework supports diversity in choosing the questions to be answered and a user-friendly framework to the users. A user-study is also conducted to judge the performance of the developed system with different background. The study shows the efficacy of the implemented system with a good level of user satisfaction over traditional CAPTCHA available today.

## I. INTRODUCTION

As more people are using Internet as a daily basis, the requirement of online services is also increasing. Many services in the internet including email, search engine, social networking are provided with free of charge. With the limited available resources, there are some cases when available services are delayed or even denied. With the expansion of web services, denial of service (DoS) attacks by malicious automated programs (e.g. web bots) is becoming a serious problem as masses of web service accounts are being illicitly obtained, bulk spam e-mails are being sent, and mass spam blogs (splogs) are being created. In order to avoid tremendous attack from malicious computer programs, CAPTCHA has been introduced to distinguish humans from computers.

Moreover, most of the online services now require users to register for identification. Most of the servers can serve a limited number of users at a given point of time. So for the consideration of performance and security, it is required to distinguish between human user and computer programs. CAPTCHA system is mainly based on the assumption that human are superior to machine to understand the images and symbols. But with the advancement of technology in text recognition and image extraction, it is now possible to extract the characters shown in CAPTCHA with satisfied accuracy. To cope up with this threat, the CAPTCHA is introduced to oblique some sequence of characters that has become really harder for a normal human being to recognize. Thus a new design concept of CAPTCHA system is a necessity.

In short, CAPTCHA (Completely Automated Public Turing Test to Tell Computers and Humans Apart) is a class of programs that is used to differentiate between human and computer programs. This classification is done through generation and grading of tests that are supposed to be solvable by only humans [4, 5]. There are some properties defined in development of CAPTCHA [15].

- Automated: Computer programs should be able to generate and grade the tests.

- Open: The underlying database(s) and algorithm(s) used to generate and grade the tests should be public. This is in accordance with the Kerckhoffs"s Principle [25].

- Usable: Humans should easily solve these tests in a reasonable amount of time. The effect of any user's language, physical location, education, and/or perceptual abilities should be minimal.

- Secure: The program generated tests should be difficult for machines to solve by using any algorithm.

There are two major issues should be considered while designing a successful CAPTCHA system: (1) robustness (difficult to break) and (2) usability (human friendly). In this paper, an idea of human cognition based CAPTCHA is presented considering above mentioned requirements. This is based on the concept that, humans can perceive the meaning of cognitive questions and answer them. But there is no such algorithm that can be used to answer those in absolute accuracy. On the subsequent section of related works proposed schemes of CAPTCHA is discussed. In the third section implementation details and the very next section survey and analysis are discussed. Before conclusion and future work result of the proposed system is discussed.

## II. RELATED WORKS

Research on CAPTCHA mechanisms has received significant attention with the aim to improve their usability and at the same time prevent adversarial attacks by malicious software. Researchers promote various CAPTCHA designs based on text and speech-recognition challenges, and image puzzle problem [1]. Nevertheless, a variety of studies have been reported that underpin the necessity for improving the usability of CAPTCHA mechanisms [4,5,6]. Result from a recent study, which investigated users' perception towards

CAPTCHA challenges; claim that current implementations do not provide an acceptable trade off solution with regards to CAPTCHA usability [2]. Another large-scale study, which evaluated CAPTCHA on the Internet's biggest websites, revealed that CAPTCHAs are difficult for humans to solve [3].

The algorithms and data used to automatically generate these CAPTCHA challenges are publicly available. But with the advancement of OCR and sophisticated image processing algorithms and tools, these text based CAPTCHAs can no longer provide the secure access to the authenticate users from malicious computer programs [16,21,24]. For instance, researchers have developed an attack against Microsoft's Hotmail CAPTCHA that yields a 60% success rate [22]. Also more complex image distortion to make it difficult for programs to crack, makes this text based method increasingly hard for human users to recognize the text, causing usability issues [23, 29].

Thus the need for new form of CAPTCHA which is automated, open, usable, and secure is of urgent need. There are some other implementations of CAPTCHA available as of now. Mostly these can be separated in three categories. Except text based scheme, there are also sound and image based CAPTCHA schemes are available. Audio based CAPTCHA was first developed for visually-impaired people [6,8]. Audio CAPTCHA usually pronounces letters or digits in randomly spaced intervals. Background noises may be added to make the tests more robust against bots. These systems are dependent on some sort of audio hardware to produce the sound clearly, and these sounds are sometimes difficult to perceive for locality reasons. Also persons with hearing difficulties cannot use this scheme. Furthermore, the basic principle to attack this CAPTCHA remains similar as text based ones, which is to extract the feature and recognize the letters. Hence, the audio based CAPTCHA scheme does not provide any more user-friendliness or robustness against bots than text based CAPTCHA [7]. In [28], a new game theorem based CAPTCHA system is proposed.

Image based CAPTCHAs inquire users to perform some forms of image recognition tasks. These systems are developed to overcome the shortcomings of previously discussed schemes of CAPTCHAs. There are some schemes that use human ability to perceive and semantically analyze images to perform a task [19]. Users are asked to categorize distorted images using noises [10] or geometric transformations [15]. There are also some methods that ask users to adjust the orientation of 3D images or to identify semantic meaning from it [13, 20]. Microsoft's Asirra [11] was designed to use the existing database of petfinder.com and prompts users to identify images of cats out of other pets. But the availability of the database and on the top of that as being it is a classification problem; Asirra is vulnerable to Machine learning attacks [12]. But content-based image retrieval and annotation techniques have shown to automatically find semantically similar images or naming them. These will allow an affordable mean of attacking image-based CAPTCHAs. User-friendliness of the systems is also a compromised factor when repeated responses are required [9] or deformed face images are shown [17,22]. Also people with color blindness have problems to figure out the distorted images. Some works are done on video CAPTCHAs [14]. Cognitive Psychology based CAPTCHA is presented in [27]. However requirement of bandwidth and difference in perception by users may be an issue.

There are some works available [18] where question-answer based CAPTCHA is shown. The author introduces the idea to use question answer method. The work in this paper is different as, it introduces the idea of different types of question rather than simple straight forward question. The authors only put the mathematical questions and hence, very few diversity in choosing the domain of the questions. Some people do not fond of answering the mathematical questions. Our proposed work also introduces a framework, where users will be given the scope to choose question group according to his/her preference. Again, as the color images are incorporated in the questions, this solution will not work for human being with color blindness. Our developed model overcomes these flaws and provides a well-accepted solution.

### III. PROPOSED MODEL

As mentioned in the related work section different types of CAPTCHA systems are used to secure web browsing. None of them gives any choice to the user to select the types of CAPTCHA. The proposed model provides the user a big window of flexibility to choose the types of CAPTCHA. By nature, humans are very responsive to answer questions. In the proposed model, a user will be provided with 5 types of CAPTCHA questions, namely, analytical, mathematical, general, text based and image based. User can select any one of the option. This system will provide 10 minutes to solve the CATPCHA problem. We want to restrict the time to prohibit the machine to analyze any single question. If enough time is given then machine can solve complex problems using artificial intelligence and pattern recognition. In that case we have to give more complex challenges to solve. It will surely affect the usability and user friendliness negatively.

Analytical type challenges provide simple analytical problems to the user. Few predefined questions are set with answers. If someone fails to give answer within the time frame another challenge is given, on the other hand if the user thinks the question is difficult to answer then try another button helps her/him to change the question instantly and reset the time. If user wants to solve mathematical problems s/he can choose mathematical option. Few simple mathematical problems are asked. The description of the mathematical questions is such that it is very easy for human being to interpret but will be very difficult to analyze by machines. User or visitor of the website has to answer within the given time limit. General type category gives the user very simple question to answer. Similar rules like analytical and mathematical types are also applicable here.

Text and image based CAPTCHA is very common now a day. Text based CAPTCHA is kept in our system because of its familiarity among users. The proposed model also provides this facility to enter text and image based CAPTCHA based on the choice from the user. For blind people, audio version of CAPTCHA is also available as an option. Analytical, mathematical and general type challenges provide the audio facility so that the blind user can choose these options.

## IV. IMPLEMENTATION

We have developed a system with few sample questions. The sample questions are for the test purposes; these should be fine grained before used in real applications. A sample scenario of the proposed model is shown in figure 1.If the user choose mathematical category a question is appear on the screen. This type of question can be answered within couple of seconds by the user but for a bots it is very difficult to answer and also take a long period of time. For example, consider the following question of mathematical category.

*"Rahim has three bananas, Karim has five apples, Sikder has seven mangos. Jamal wants to buy three apples. How many apples left to karim?"*

It is clear from the above mentioned question that the user can give the correct answer easily but it is very difficult for a machine.

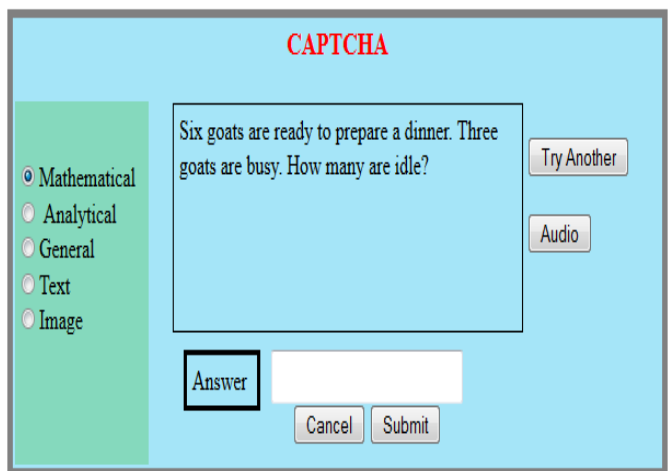

Fig. 1. Proposed CATCHA System

System receives the answer form the user and compares it with the answer stored in the database. User can give the answer in any case (e.g., uppercase or lowercase), before comparing with the database the system makes all answers in lower case. For the incorrect answer error message is shown and it asks to try another one or change the category. Next option is given based on the user's choice. Table 1 shows few sample questions of each category.

## V. SURVEY AND ANALYSIS

### A. Initialize the survey

To check the acceptance level and usability of the proposed model a session has been organized with 100 students and teachers from different departments of Daffodil International University. Along with 10 teachers, 20 Students from Computer Science and Engineering department, 15 from Electrical and Electronics Engineering department, 15 from Software Engineering department participated in the session. Other 40 students are from English, law, BBA and journalism departments. The main focus of this session is to collect the user's opinion about the newly designed framework.

First the new framework is given to the participants to explore without giving any instructions. After playing with the system a set of questions is given to them and asked to give their feedback. Each question has four options like strongly agree, agree, partially agree and disagree.

TABLE I. SAMPLE QUESTIONS OF EACH CATEGORY

| Category | Questions |
|---|---|
| Analytical | Mina had orange and mango. Mina ate orange. Which fruit is left? |
| Mathematical | Karim's age is one third to his father. His father age is 45. How old karim is? |
| General | In which direction does the Sun rise? |
| Text | ![28ivw] |
| Image | ![lion] |

## VI. RESULT ANALYSIS

As discussed above out of 100 participants 50 are from technical background and 50 from non technical background. Responses from both groups are collected separately for better justification of the system. Concerns of the female participants are also taken separately. Following charts shows the impact of our system. Question by question analysis is given below.

### B. Technical Perspective

Firstly, we will analyze the survey results from the technical perspective. We will try to find if there is any significant difference between technical and non-technical people. We are interested about the non-technical people because in today's digital world is dominated by non-technical people.

First question was asked to the participants about whether the new framework is easy to use or not? As shows in the figure 2, 66% technical knowledge based participants strongly agree that the new framework is easy to use whereas 70% general participants strongly agree with that, 28% technical people agree in contrast with 20% non technical. 4% partially agree where none from non technical partially agree with that. Rest 2% technical people said it is not easy to use where it is 10% from non general people. So, from this perspective we can conclude that although non technical people are more in favor of this kind of system but some of them are against of this new system whereas all the technical people are in support of the new system.

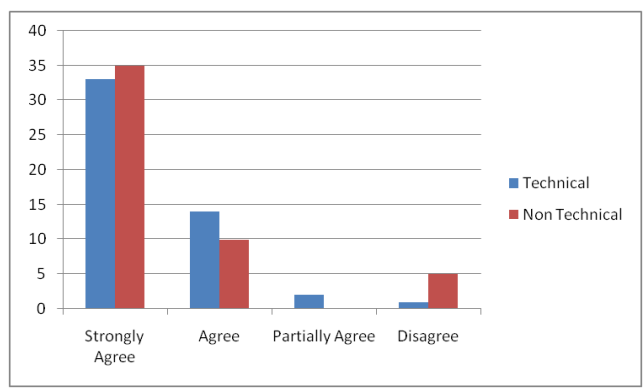

Fig. 2. : The new framework is easy to use or not

Second question was asked about the probability of making mistakes compared to existing CAPTCHA system (reCaptcha [26]). 70% of technical and 80% of non technical people think that the new system has less chance to make any error. 22% and 10% also agree. 4% of both technical and non technicalsurveyee shows their disagreement. Figure 3 reflects the opinion.

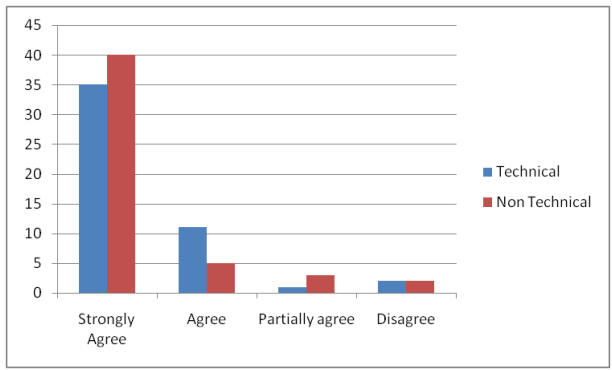

Fig. 3.   This system has less chance to make error

Third question was asked about the brain work to solve any given problem. 96% participants from technical group agreed that brain work is needed to solve the given problem whereas 20% participants from non technical group agree. 60% non technical people strongly agree with that the system need brain work. Rests of the participants think differently. Figure 4 shows the result.

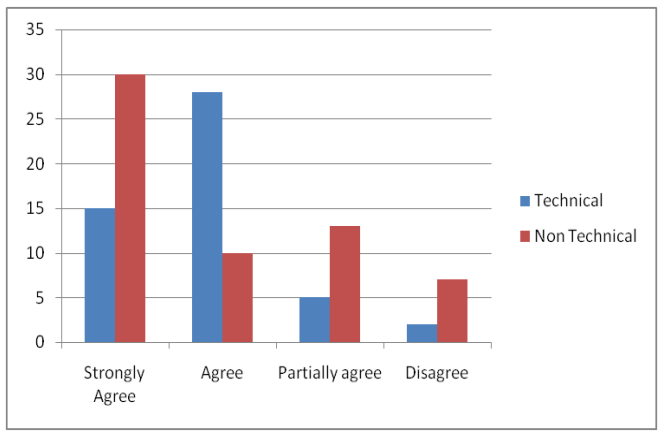

Fig. 4.   Do you need any brain work to solve the problem given?

Fourth question was asked to the participants that the system is user friendly and helpful or not? From figure 5 it is clear that almost similar percentage participants from both technical and non technical group strongly agreed that the new system is very much helpful as well as user friendly. No one express their dissatisfaction. This shows strong support for the new proposed system.

Fifth Question was whether the system is better for all types of user or not? Figure 6 gives a clear idea that 80% and 90% form technical and non-technical group respectively strongly agree that the new system is better for all types of user. No one express their dissatisfaction.

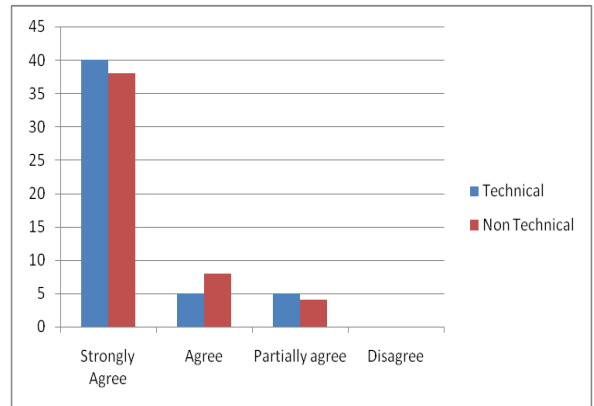

Fig. 5.   The system is user friendly and helpful?

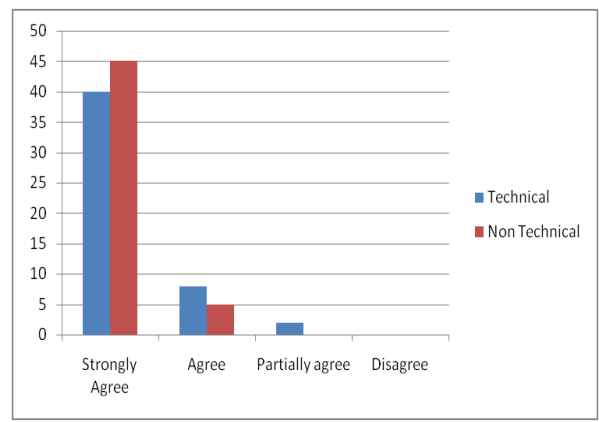

Fig. 6.   System is better for all types of user or not?

Sixth and the last question was to check the satisfaction level of the user. 68% participants from technical group and 94% from non technical group strongly agree that they are satisfied with the system. 24% are agreeing on that. 6% and 4% are partially agreed. Rests of them are disagreed. Figure 7 shows the proof.

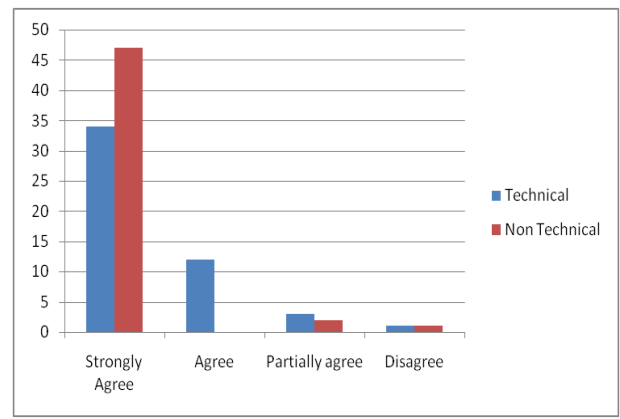

Fig. 7.   Are you satisfied with the system?

So from the analysis of the technical perspective we can conclude that both technical and non-technical people are in strongly in favor of the proposed system. Specially, non-technical people are more in favor of such system.

*C. Gender perspective*

We have also investigated the gender factor in CAPTCHA system. 20 females have participated in the survey where 10 of them are from non-technical background. Almost 95% from non technical group express their deep satisfaction about the proposed system. More than 80% from technical group has given their consent on satisfaction. Based on the data collected form female participants it is clear that they are biased on the design and easiness of the system where as the male participants and also technical group people concern about the effectiveness of the system. Two questions were selected to test the satisfaction level of female participants.

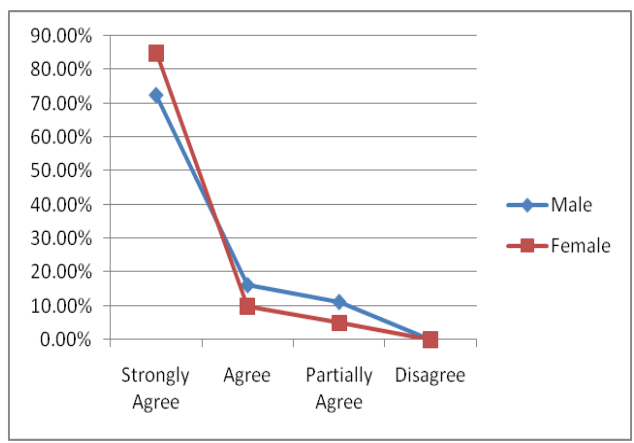

Fig. 8. System is user friendly and helpful

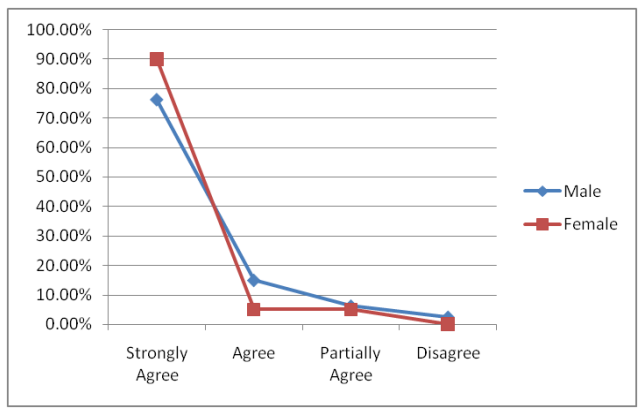

Fig. 9. Are you satisfied with the system?

From the above two figures (Figure 8 and figure 9) it is clear that female participants are more positive about the satisfaction, friendliness and usefulness of the system rather than male participants.

*D. Time Perspective*

We have also analyzed the time required to answer each type of question. From Table 2, we can easily see that Text based question (current reCAPTCHA) need more time compared to any other types.

The response is subject to the type of questions and may vary slightly for different set of questions.

TABLE II. TIME TAKEN TO ANSWER THE QUESTIONS

|  | Q # 1 | Q # 2 | Q # 3 | Q #4 | Q # 5 | Average time (in seconds) |
|---|---|---|---|---|---|---|
| Mathematical | 7.99 | 9.67 | 7.51 | 6.73 | 6.48 | 7.76 |
| Analytical | 5.33 | 4.63 | 7.54 | 6.12 | 4.25 | 5.57 |
| General | 2.53 | 2.45 | 4.89 | 3.24 | 3.15 | 3.25 |
| Text | 9.36 | 8.51 | 11.87 | 7.95 | 10.82 | 9.70 |
| Image | 4.87 | 4.62 | 5.43 | 5.98 | 5.33 | 5.24 |

*  Q#1 means question 1 and so on

*E. Likeliness to attempt any category*

We have identified the likeness of the user for any category of questions. From our survey it has revealed that most of the people like the general questions and least number of people attempted the text based system. Figure 10 reflects the likeliness of the participants.

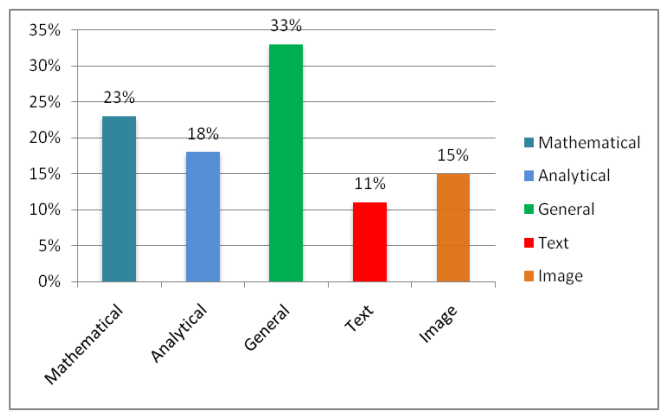

Fig. 10. Likeliness of Participants to attempt any Category

Table 3 shows the mode and means value of the selected 6 questions. The result is also categorized in technical and non-technical background to cover the diversity of the participants of the users. With analyzing the mode value for all participant data, it can be derived that most people strongly agreed on the given set of questions. Hence, it implies that the proposed model is accepted by majority of the participants. The mean value also justify this same results.

From technical perspective it is clear that the mean and mode value of all question asked in the survey is near about 3. That indicates the high acceptance and effectiveness of newly proposed system among technical people. For the non-technical group, the mean values of responses are greater than technical people all the questions. This clearly indicates the efficacy and usability of the framework among majority of the common people. The questions in the questionnaire are designed to reflect 3 criteria (Learn ability, Efficiency and Satisfaction) of usability study.

TABLE III. MEAN AND MODE OF REPLIES OF THE TECHNICAL AND NON-TECHNICAL SURVEYED.

| Question number | For Technical participants | | For non Technical participants | |
|---|---|---|---|---|
| | Mean | Mode | Mean | Mode |
| 1 | 2.48 | 3 | 2.4 | 3 |
| 2 | 2.28 | 3 | 2.76 | 3 |
| 3 | 1.51 | 3 | 2.32 | 3 |
| 4 | 2.45 | 3 | 2.7 | 3 |
| 5 | 2.51 | 3 | 2.9 | 3 |
| 6 | 2.28 | 3 | 2.86 | 3 |

## VII. LIMITATION

Though system has the option to listen audio for disabled person but none of them are found while taking the survey. So audio based CAPTCHA is not tested in this survey.

## VIII. CONCLUSION AND FUTURE WORK

This paper illustrates a new design for CAPTCHA system based on human cognition. This model demonstrates the ability of human to find the answer that other bots and external programs fail to interpret and evaluate. The conducted survey explains the usability of this new form of CAPTCHA and provides valuable feedback to design the overall system and types of question pattern. This framework can easily be extended to specific website to include question of any particular area of interest. In future, more extensive user-study will be performed to suggest context aware questions to give the most user-friendly experience in web surfing and also combat against CAPTCHA farming.

## ACKNOWLEDGEMENT

We are expressing our heartfelt thanks to all of the participants specially the students of Daffodil International University.